\documentclass[twocolumn,apj]{emulateapj}
\usepackage{color}

\def\mbf#1{\mbox{\boldmath ${#1}$}}
\def\Alfven{Alfv\'{e}n~}

\slugcomment{Submitted to ApJ}

\shorttitle{Disk Wind Driven by MRI}
\shortauthors{Suzuki \& Inutsuka}

\begin{document}

\title{Disk Winds Driven by Magnetorotational Instability and 
Dispersal of Proto-Planetary Disks}

\author{Takeru K. Suzuki$^{1}$ \& Shu-ichiro Inutsuka$^{2}$}
\email{stakeru@ea.c.u-tokyo.ac.jp}
\altaffiltext{1}{School of Arts and Sciences, University of Tokyo,
Komaba, Meguro, Tokyo, Japan, 153-8902}
\altaffiltext{2}{Department of Physics, Kyoto University, 
Kyoto, Japan, 606-8502}

\begin{abstract}
By performing local three-dimensional MHD 
simulations of stratified accretion disks, 
we investigate disk winds driven by MHD turbulence. 
Initially weak vertical magnetic fields are 
effectively amplified by magnetorotational instability and 
winding due to differential rotation. 
Large-scale channel flows develop most effectively at 1.5 - 2 times 
the scale heights where the magnetic pressure 
is comparable to but slightly smaller than 
the gas pressure. 
The breakup of these channel flows drives structured 
disk winds by transporting the Poynting flux to the gas.  
These features are universally observed in the simulations of 
various initial fields. 
This disk wind process should play an essential role in the 
dynamical evaporation of protoplanetary disks. 
The breakup of channel flows also excites the momentum fluxes associated with 
Alfv\'{e}nic and (magneto-)sonic waves toward the midplane, which 
possibly contribute to the sedimentation of small dust grains 
in protoplanetary disks.  
\end{abstract}

\section{Introduction}
Magnetorotational instability \citep[MRI;][]{bh91} is regarded as 
a robust mechanism to provide turbulence for an efficient outward transport 
of angular momentum in accretion disks. 
MHD simulations in a local shearing box have been carried out 
\citep[e.g.,][]{hgb95,bra95,san04} to 
study the properties of MRI-driven turbulence. \citet{ms00} studied vertically 
stratified local disks with free boundaries to allow 
leaks of mass and magnetic field. While their main purpose is to study general 
properties of stratified disks such as disk coronae rather than disk winds, 
they concluded that the mass flux of the outflows is small in the cases of 
initially toroidal and zero-net vertical flux magnetic fields. 
 
On the other hand, 
protoplanetary disks around young stars should have 
net vertical magnetic fields that are connected to 
their parental molecular clouds. 
In this case, the physical conditions of the surface of the disk 
are analogous to the open coronal holes of the Sun 
where the solar wind is driven by turbulent footpoint motions 
of the magnetic field lines \citep{sak07,tsu08}. 
Obviously, MHD turbulence excited by MRI in the disk 
is also expected to drive winds from the surfaces of the accretion disk. 
Although such a disk wind mechanism may play a significant 
role in the evolution of accretion disks \citep{fdc06}, 
quantitative studies have not been carried out so far 
because of difficulties in the numerical treatment: 
a long-term calculation of the wind process requires 
accurate description of outgoing boundary conditions for 
various types of waves, rather than simple free boundaries 
\footnote{Simple free boundaries means setting the derivatives of variables 
 to be zero. 
 In this case, however, unphysical reflections of waves usually occur. 
 For the real outgoing boundary, only the derivatives of incoming 
 characteristics should vanish \citep{tho87}.}.     


In this Letter, we investigate disk winds driven by MRI with 
initially vertical magnetic fields utilizing the rigorous outgoing 
boundary condition that was originally developed for the simulations 
of solar \citep{si05,si06} and stellar \citep{suz07} winds.

\section{Setup}

We perform 3D MHD simulations in a local shearing box (Hawley et al.1995), 
taking into account vertical stratification \citep{stn96,ts07}.  
We set the $x$-, $y$-, and $z$-coordinates as the radial, azimuthal, 
and vertical directions, respectively. 
We solve the ideal MHD equations with an isothermal equation of state 
in a frame corotating with Kepler rotation. 
In the momentum equation we consider the vertical gravity by a central star, 
$\Omega_0^2\mbf{z}$, where $\Omega_0$ is Keplerian rotation frequency.
We adopt a second-order Godunov-CMoCCT scheme, in which we solve nonlinear 
Riemann problems with magnetic pressure at cell boundaries for compressive 
waves and adopt the consistent method of characteristics (CMoC) for the 
evolution of magnetic fields \citep{cl96}. 

The simulation region is $(x,y,z)=(\pm 0.5H_0,\pm 2H_0,\pm 4H_0)$, and is 
resolved by (32,64,256) grid points, where $H_0=\sqrt{2}c_s/\Omega_0$ is 
the pressure scale height
for sound speed, $c_s$.  
The shearing boundary is adopted for the $x$-direction 
to consider the Keplerian shear flow (Hawley et al. 1995). 
The simple periodic boundary is adopted for the $y$-direction. 
We prescribe the outflow condition in the $z$-directions 
by adopting only outgoing characteristics from all seven (six for 
isothermal gas) MHD characteristics at the $z=\pm 4H_0$ boundaries 
(Suzuki \& Inutsuka 2006). 
While the $z$-component of the magnetic flux is strictly conserved, 
the $x$- and $y$- components are not conserved because of 
the outgoing condition at the $z$ boundaries.   
We initially set up a hydrostatic density structure, 
$\rho=\rho_0\exp(-z^2/H_0^2)$, with $\rho_0=1$,  and a constant 
vertical magnetic field, $B_{z,0}$, with the plasma $\beta$ value, 
$\beta_0=8\pi \rho_{0}c_s^2/B_{z,0}^2=10^6$ (for our fiducial model) 
at the midplane. We use $\Omega_0=1$ and 
$H_0=1$, which gives $c_s^2=1/2$.  
Small random perturbations, $\delta v=0.005 c_s$, are initially given for 
the seeds of MRI. 

\section{Results}
\begin{figure*}[h]
\figurenum{1} 
\epsscale{1.}
\begin{center}
\plotone{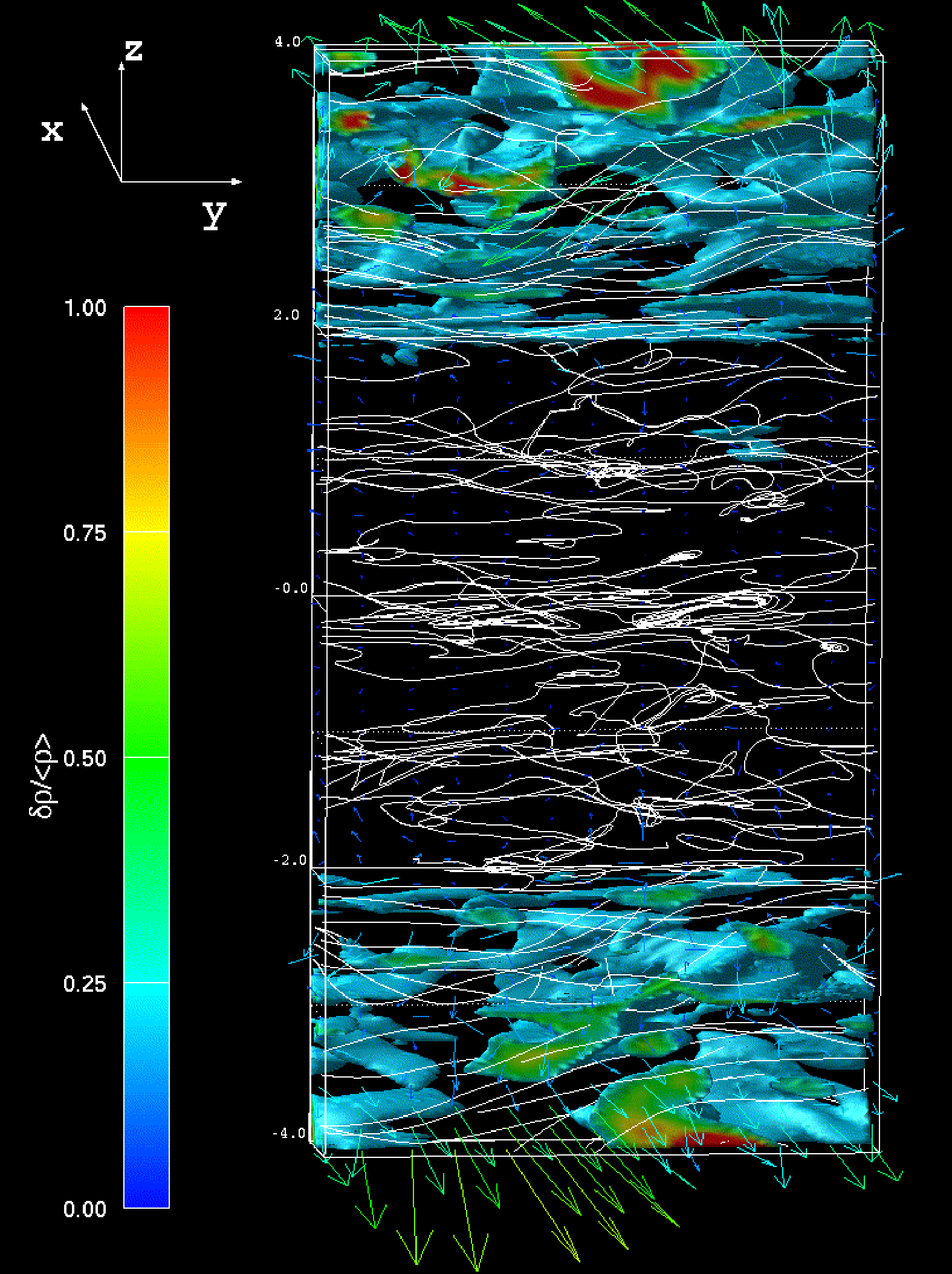}
\end{center}
\caption{Snapshot of the local disk structure at $t=210$ rotations. 
The white solid lines are magnetic fields, the arrows indicate velocity 
 fields, and 
 the colored region corresponds to $\delta\rho/{\langle\rho\rangle}>0$.
}
\label{fig:snp}
\end{figure*}

In most of the simulation region, 
the initial magnetic fields are moderately weak, so it is unstable 
with respect to MRI.  
In $|z|<2.5H_0$, however, we cannot initially
resolve the wavelengths, $\lambda_{\rm max}\approx 2\pi v_{\rm A}/\Omega_0$, 
of the most unstable mode because 
$\lambda_{\rm max}<\Delta z(=H_0/32)$, where $v_{\rm A}=B/\sqrt{4\pi\rho}$ 
is \Alfven speed and $\Delta z$ is the mesh size 
(the initial $\lambda_{\rm max}\approx 0.2\Delta z$ at the midplane).
First, MRI develops around $z\approx\pm 3H_0$ after $\approx 3$ 
rotations. 
The turbulence driven by MRI gradually spreads toward the 
midplane, since the growth time (approximately 
$\propto\Delta z/\lambda_{\rm max}$ for $\Delta z>\lambda_{\rm max}$)
of the resolved wavelength is longer there.
When $t\gtrsim 100$ rotations, the midplane finally becomes turbulent. 
The magnetic field 
strength saturates in the entire box after $t\gtrsim 200$ rotations, and the 
system becomes quasi-steady-state. 
At this time, $\lambda_{\rm max}$ can be resolved even at the midplane owing 
to the increase of the field strength. 
We continue the simulation further up to 400 rotations.  

Figure \ref{fig:snp} is the snapshot of 
magnetic field (white lines), velocity field (arrows), and 
$\delta\rho/{\langle\rho\rangle}$ (color) at $t=210$ rotations, where 
${\langle\rho\rangle}$ is the density 
averaging over each $x$-$y$ plane and $\delta\rho=\rho-\langle\rho\rangle$.
The magnetic fields are turbulent, dominated by the toroidal ($y$) 
component because of winding. Angular momentum is outwardly transported 
by anisotropic stress due to the MHD turbulence. 
At the saturated state, 
$\alpha\equiv(v_{x}\delta v_{y}-\frac{B_{x}B_{y}}{4\pi\rho})/c_s^2$ 
is $\sim 0.01$ in the midplane. 
One can also observe that the structured outflows 
stream out from both the upper and lower boundaries.
Below we inspect the properties of the outflows in more detail.


\begin{figure}
\figurenum{2} 
\epsscale{1.1}
\begin{center}
\plotone{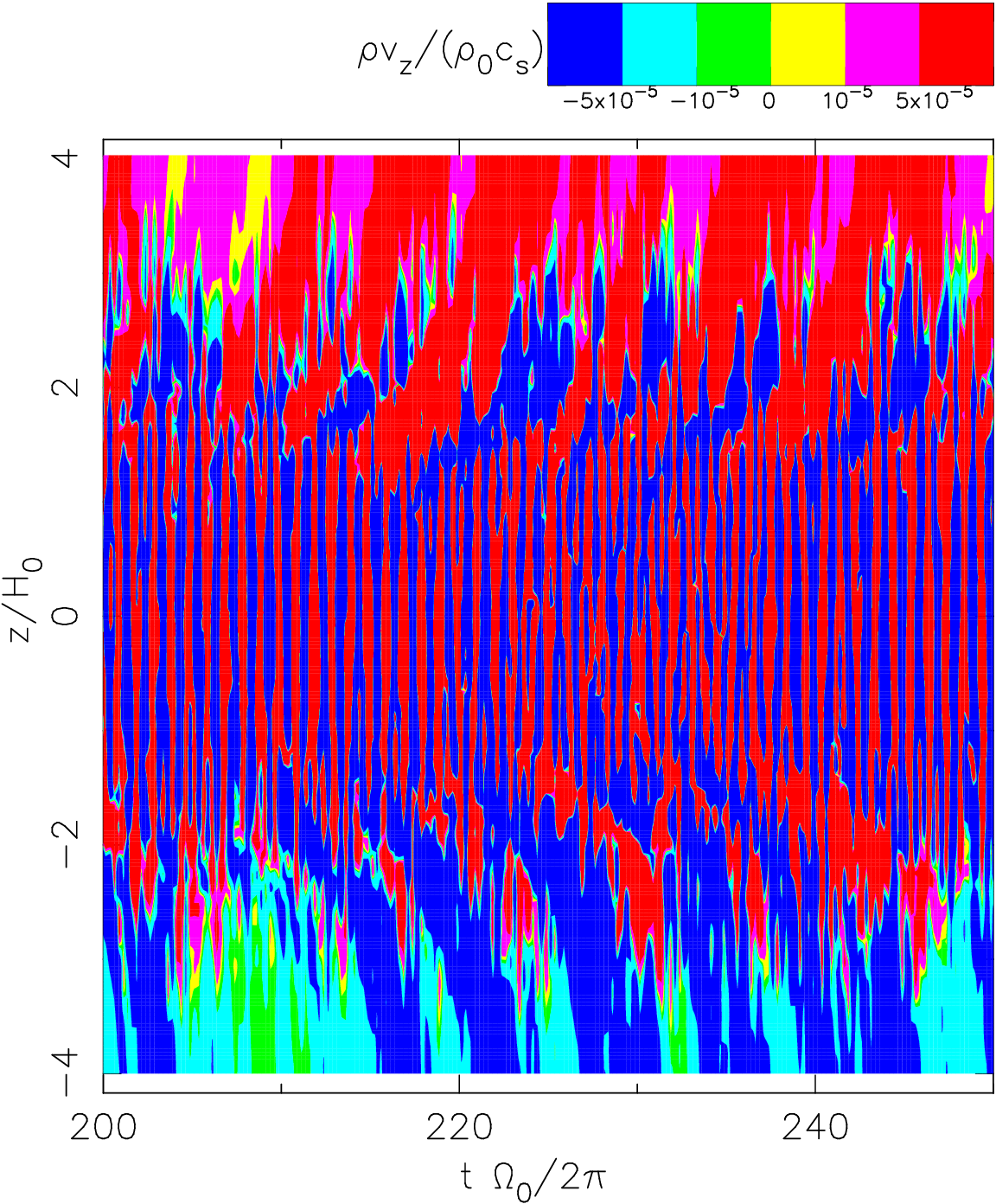}
\end{center}
\caption{Time-height diagram of the mass flux, $\rho v_z$, 
normalized by $\rho_{0}c_s$. 
$\rho v_z$ is averaged on the $x$-$y$ plane at each height, $z$ 
(vertical axis).
The unit of of horizontal axis is the rotation period ($2\pi/\Omega_0$).}
\label{fig:tds}
\end{figure}

Figure \ref{fig:tds} presents the mass flux of the $z$-component, 
$\rho v_z/\rho_{0}c_s$, in the $t-z$ plane. 
We averaged $\rho v_z$ on the $x$-$y$ plane at each $z$ grid point. 
One can see that the gas flows out from both the upper and lower boundaries. 
The mass fluxes near the surface regions are highly time dependent with a 
quasi-periodic cycle of $\sim 5-10$ rotations. Moreover, from $z\sim\pm 2H_0$ 
the mass fluxes direct to the midplane, almost coinciding with the 
periodicities of the outflow fluxes. In other words, the mass flows are 
ejected to both upward and downward directions from `injection regions' 
located at $z\sim\pm 2H_0$.

These features are consequences of the breakup of channel flows 
\citep[e.g.,][]{si01,san04}. 
At $z\sim\pm 2H_0$, the wavelength of the most unstable mode 
with respect to MRI, $\lambda_{\rm max}$, is comparable with the scale height, 
$H_0$. In the region $|z|>2H_0$, $\lambda_{\rm max}>H_0$; hence, 
it is stable against MRI. 
In the region $|z|<2H_0$, smaller-scale turbulence develops preferentially 
because $\lambda_{\rm max}<H_0$.   
Therefore, at $z\sim\pm 2H_0$ the largest scale channel flows 
 develop, and their breakup by reconnections\footnote{
 We do not explicitly include the physical resistivity term in the calculation 
 shown in this Letter, and so, the reconnections are due to 
 the numerical effect determined by the grid scale.} 
drives the mass flows to both upward and downward directions. 
In the region $|z|<2H_0$, 
the gas pressure largely dominates the magnetic pressure 
so that strong mass flows cannot be driven by the magnetic force associated 
with reconnections between small-scale turbulent fields.  
The periodic oscillation 
of 1 Keplerian rotation time
around the midplane is the vertical (epicycle) motion. 

\begin{figure}[b]
\figurenum{3} 
\epsscale{1.22}
\begin{center}
\plotone{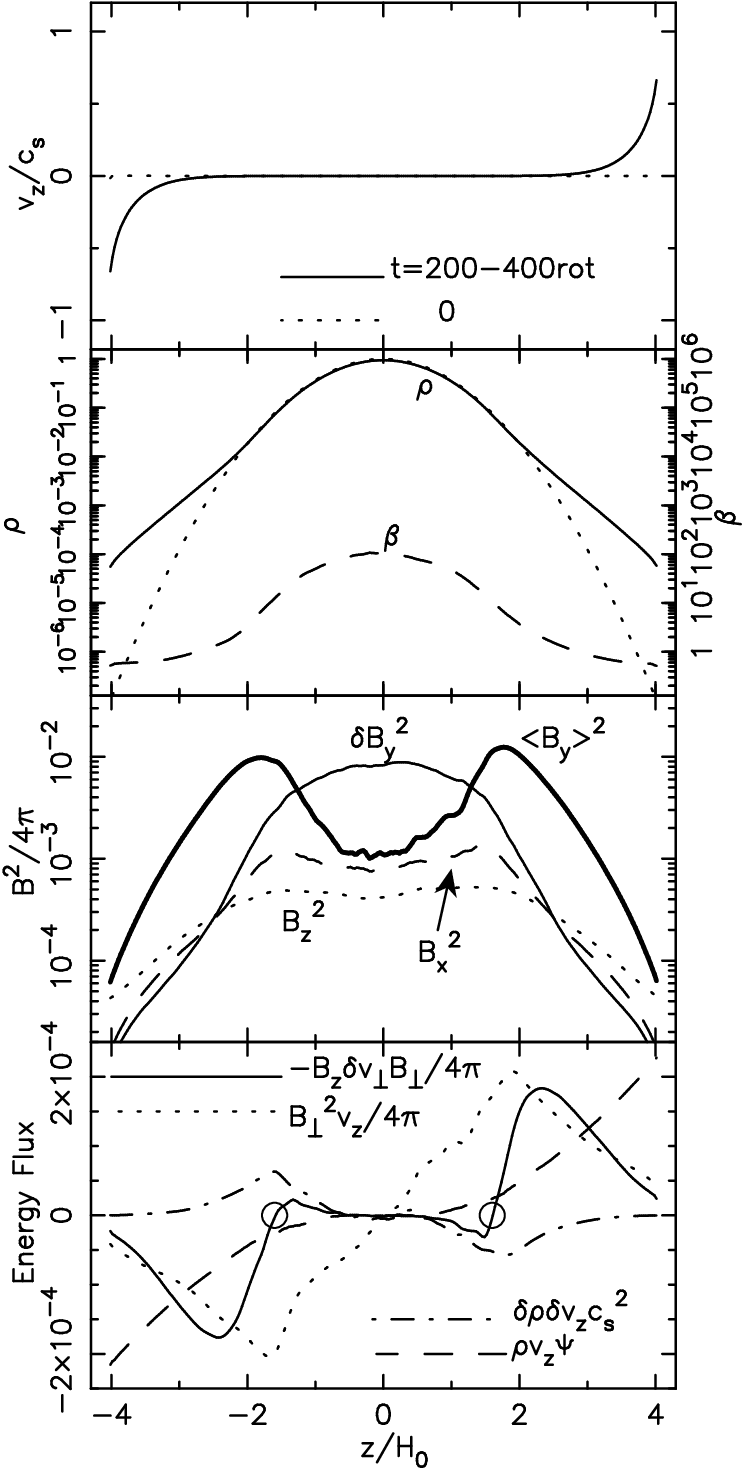}
\end{center}
\caption{Time-averaged disk structure during $t=200$-$400$ rotations. 
The variables are also averaged on the $x$-$y$ plane at each $z$ grid. 
The top panel shows $v_z/c_s$ (solid), whereas the dotted line is the initial 
condition ($v_z/c_s=0$). The second panel presents density (solid; left axis) 
and plasma $\beta$ (dashed; right axis), in comparison with the initial 
condition (dotted; for both density and plasma $\beta$). The third panel 
presents the magnetic energy, $B^2/4\pi$. The dashed, solid, and dotted lines 
correspond to the $x$-, $y$-, and $z$-components, and the $y$-component 
shows both mean (thick) and fluctuation (thin) components. The bottom 
panel illustrates the energy flux in units of $\rho_0(H_0\Omega)^3$. 
The solid and dotted lines are the Poynting flux 
associated with the magnetic tension 
($-B_z\delta v_{\perp}B_{\perp}/4\pi$) and the magnetic energy 
($B_{\perp}^{2}v_z/4\pi$). The dot-dashed line is the net energy 
flux due to sound waves ($\delta\rho\delta v_{z}c_{s}^2$: see the text). 
The dashed line is the term concerning the potential energy ($\rho v_z\Phi$). 
The circles are the `injection regions' defined as the locations where the 
signs of $-B_z\delta v_{\perp}B_{\perp}/4\pi$ change. }
\label{fig:tav}
\end{figure}

Figure \ref{fig:tav} presents the disk wind structure averaged over 200 - 400 
rotations. The variables are averaged on the $x$-$y$ plane at each $z$ point. 
The top panel shows that the average outflow velocity is nearly the sound speed 
at the upper and lower surfaces. 
The second panel presents the structures of density and plasma $\beta$ value. 
The comparison of the final density structure (solid) with the initial 
hydrostatic structure (dotted) shows that the mass is loaded up to the onset 
regions of outflows from $z\approx\pm 2H_0$. 
In the wind region $|z|\gtrsim 3H_0$, $\beta$ is below unity;
the disk winds start to accelerate when the magnetic 
pressure dominates the gas pressure. 
    
The third panel shows magnetic energy at the saturated state. 
The dashed, solid, and dotted lines are $x$-, $y$-, and $z$-components, 
respectively. 
In the $y$-component we show both mean, $\langle B_y\rangle^2$ and 
fluctuation, $\delta B_y^2$, components. $\langle B_y\rangle^2$ 
is the simple average on the $x$-$y$ planes, $\langle B_y(z)\rangle=\int\int 
dx dy B_y(x,y,z)/(L_{x}L_{y})$, and the fluctuations are 
determined from $\delta B_y^2(z)=\int\int dx dy(B_y(x,y,z)-\langle 
B_y(z)\rangle)^2/(L_{x}L_{y})$, where $L_x(=H_0)$ and 
$L_y(=4H_0)$ are the $x$ and $y$ lengths of the simulation box. 
As for $B_x$ and $B_z$ 
the fluctuation components greatly dominate the means.  
The magnetic energy, which is dominated by the toroidal ($y$) component 
as a consequence of winding, is amplified by $\approx$ 1000 times of
the initial value ($B_{z,0}^2/4\pi=10^{-6}$) in most of the region 
($|z|<3H_0$). 
While in the region near the midplane ($|z|<1.5H_0$), 
the magnetic field is dominated by the fluctuating component 
($\delta B_y$), the mean component dominates in the regions near the 
surfaces ($|z|>1.5H_0$). 
In the surface regions the magnetic pressure 
is comparable to or larger than the gas pressure ($\beta\lesssim 1$), 
and so, the gas motions cannot control the configuration of 
the magnetic fields. 
Therefore, the field lines tend to be straightened by magnetic tension 
to give $\langle B\rangle^2>\delta B^2$ there, 
even if the gas is turbulent.  
We also note that $\langle B_z^2\rangle $ 
 is amplified by MRI and Parker (1966) instability, 
 whereas $\langle B_z\rangle^2$ is strictly conserved.

The bottom panel shows the status of energy transfer.   
The $z$-component of the total energy flux 
is expressed as 
\begin{equation}
\rho v_z\left(\frac{1}{2} v^2+\Phi+h\right)+v_z\frac{B_{\perp}^2}{4\pi}-\frac{B_z}{4\pi}(v_{\perp}B_{\perp}),
\label{eq:efx}
\end{equation}
where $h$ is the enthalpy\footnote{Formally, $h=c_s^2\log\rho$ for isothermal 
gas}, 
$\Phi=z^2/2$, and $\perp$ denotes $x$ and $y$; 
e.g. $B_{\perp}^2=B_x^2+B_y^2$. 
The Poynting flux is separated from 
the term of direct transport of magnetic energy 
($B_{\perp}^{2}v_z/4\pi$) and the term related to 
magnetic tension ($-B_{z}v_{\perp}B_{\perp}/4\pi$). 
The solid and dotted lines are respectively 
$-B_z\delta v_{\perp}B_{\perp}/4\pi$ and $B_{\perp}^{2}v_z/4\pi$, 
where $\delta v_{\perp}=v_x$ 
and $v_y+3/2\Omega_{0}x$ ($-3/2\Omega_{0}x$ is the background Kepler rotation). 
The dashed line is the potential energy term, $\rho v_z\Phi$. 
The dot-dashed line is the energy flux of sound waves, 
$\delta\rho\delta v_{z}c_s^2$ 
(see below).  
The gas pressure ($pv_z$) and hydrodynamical 
turbulent pressure ($\rho\delta v^{2}v_z/2$) are smaller than 
these terms.  
The kinetic energy flux of the winds ($\frac{1}{2}\rho v_z^3$) is 
also small $\lesssim 10^{-5}$ at the outer boundaries.  

The figure shows that the materials near the surfaces are lifted up by the 
conversion of the Poynting flux; 
the absolute values of the Poynting flux terms (solid and dotted) decrease
with height 
in the region $|z|\gtrsim 2H_0$, and the absolute value of the 
potential energy flux (dashed) increases. 
Both magnetic pressure and tension terms contribute almost equally.    

$-B_z\delta v_{\perp}B_{\perp}/4\pi$ 
and $\delta\rho\delta v_{z}c_{s}^2$ are the net energy fluxes of \Alfven waves 
and sound waves to the $+z$-direction. 
$-B_z\delta v_{\perp}B_{\perp}/4\pi$ can be rewritten as
\begin{equation}
-\frac{1}{4\pi}B_z\delta v_{\perp}B_{\perp}=\rho v_{{\rm A},z}(\delta 
v_{\perp,+}^2-\delta v_{\perp,-}^2),
\end{equation}
where $v_{{\rm A},z}=B_z/\sqrt{4\pi\rho}$, 
and $\delta v_{\perp,\pm}=\frac{1}{2}(\delta v_{\perp}\mp 
B_{\perp}/\sqrt{4\pi\rho})$ are 
Els\"{a}sser variables, which correspond to the amplitudes of \Alfven waves 
propagating to the $\pm z$-directions.  
$\delta\rho\delta v_z c_s^2$ is also rewritten as 
\begin{equation}
\delta\rho\delta v_z c_s^2=\rho c_s(\delta v_{\parallel,+}^2-\delta 
v_{\parallel,-}^2 ),
\end{equation}
where $\delta v_{\parallel,\pm}=\frac{1}{2}(\delta v_z\pm 
c_s\frac{\delta\rho}{\rho})$ denote the amplitudes of sound 
waves\footnote{Strictly speaking, these are magnetosonic waves, 
namely the fast mode in the high $\beta$ plasma, and the slow 
mode that propagates along $z$ in the low $\beta$ plasma. 
Note also that the signs are opposite for $\delta v_{\perp,\pm}$ and 
$\delta v_{\parallel,\pm}$, reflecting the transverse and longitudinal 
characters.} propagating in the $\pm z$-directions. 

An interesting feature of $-B_z\delta v_{\perp}B_{\perp}/4\pi$ is that the 
sign changes at $z\approx\pm 1.6H_0$ (the circles in the bottom panel); 
in $z>1.6H_0(<-1.6H_0)$, the flux is upward (downward) and $|z|<1.6H_0$ 
the flux is toward the midplane. 
This is a consequence of the breakup of channel flows, 
as described previously. 
Reconnections break up channel flows and 
generate large amplitude ($|\delta{B_\perp}|>|B_z|$) Alfv\'{e}n(ic) waves in 
both upward and downward directions. 
$\delta\rho\delta v_z c_s^2$ is also directed to the midplane, namely 
sound(-like) waves are generated by the reconnections. The absolute 
value of the energy flux peaks at the injection regions at 
$z\approx\pm 1.6H_0$.
The energy flux of sound waves to the midplane is larger than that of 
the \Alfven waves. 
On the other hand, 
the \Alfven wave component greatly dominates 
in the flux to the surfaces. 
This is because $\beta>1$ (the gas pressure 
dominates) in $|z|\lesssim 2H_0$ and $\beta\lesssim 1$ near the surfaces. 
 



\begin{figure}[h]
\figurenum{4} 
\epsscale{0.9}
\begin{center}
\plotone{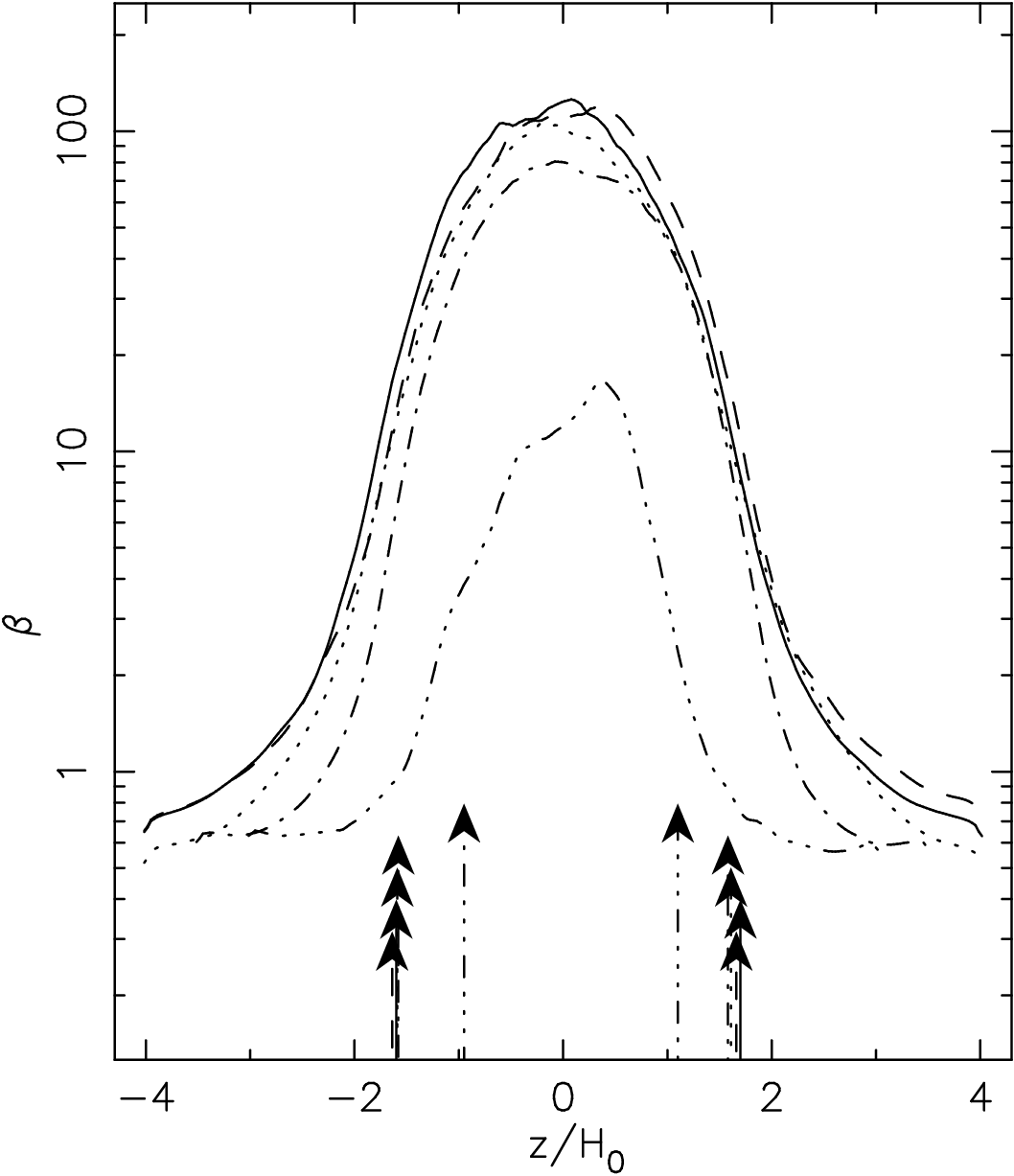}
\end{center}
\caption{Dependence of the final $\beta$ structure on initial 
magnetic fields. 
The dash-dot-dotted, dot-dashed, dotted, solid, and dashed lines correspond to
the initial vertical fields with $\beta_0=10^4,10^5,10^6$, and $10^7$, 
and the initially toroidal field cases. 
The arrows indicate the locations of the injection regions defined as 
the points where the sings of $-B_z\delta{v}_{\perp}B_{\perp}$ change. 
Longer arrows correspond to smaller $\beta_0$ (the shortest arrows are for the 
initially toroidal case). 
}
\label{fig:betadep}
\end{figure}

\begin{figure}[h]
\figurenum{5} 
\epsscale{0.9}
\begin{center}
\plotone{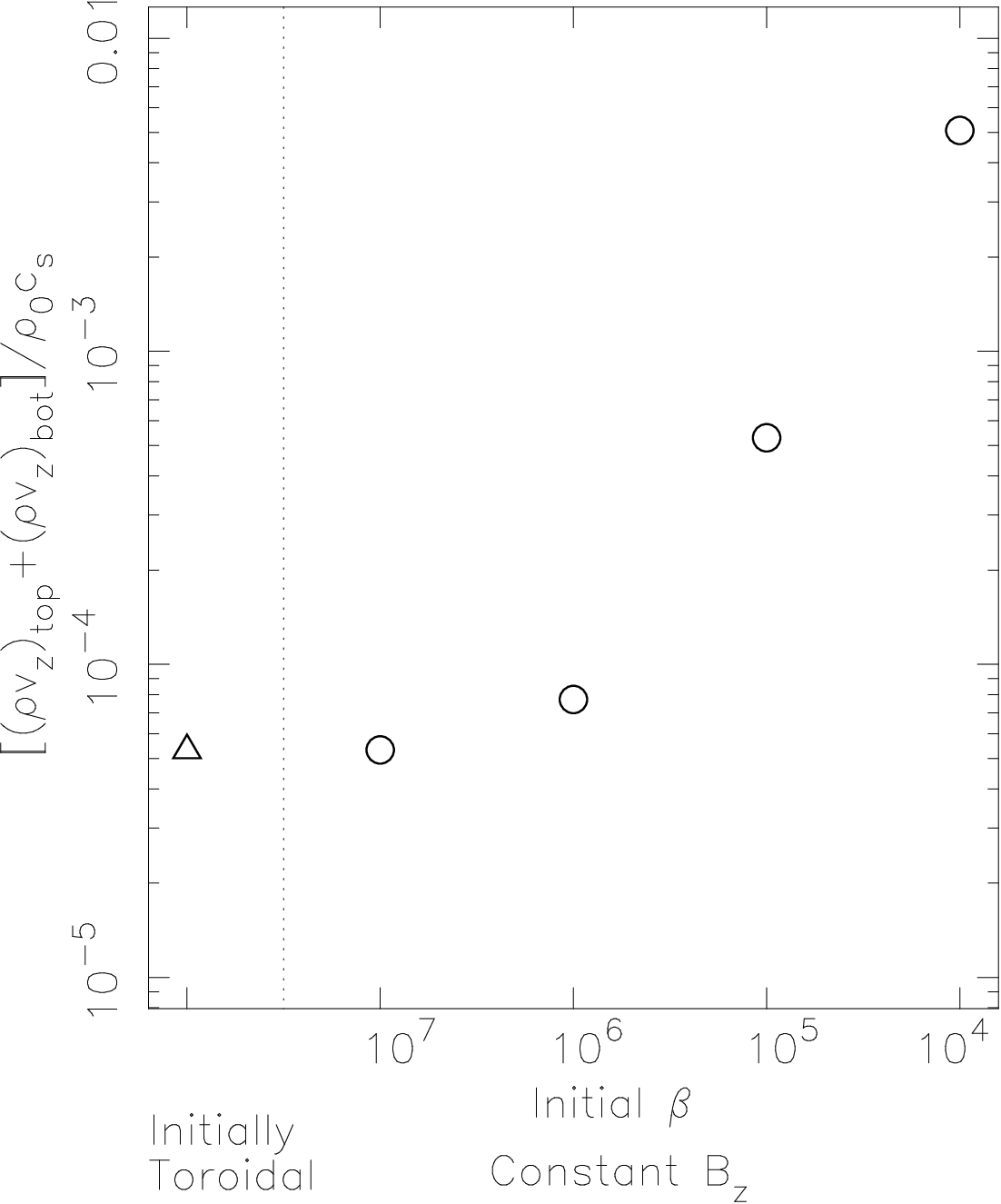}
\end{center}
\caption{Sum of the mass fluxes normalized by 
$\rho_{0}c_s$ of the disk 
winds from the top and bottom boundaries of the simulation box.
The horizontal axis indicates the initial $\beta_0$ at the midplane for the
vertical field cases {circles}. 
The initially toroidal case is plotted at the leftmost location (triangle).}
\label{fig:msfxdep}
\end{figure}

In order to study the effects of the initial magnetic fields, we 
performed the simulations with initial different vertical field strengths, 
$\beta_0=10^4,10^5,10^6,10^7$, and initially toroidal field in $|z|<3H_0$ 
with $\beta_0=10^6$ at the midplanes. 
Figure \ref{fig:betadep} compares $\langle\beta\rangle$ structures. 
Figure \ref{fig:msfxdep} compares 
 the sum of disk mass fluxes from the upper and lower boundaries 
 with various cases for the initial $\beta_0$ at the midplane.
In all cases except the $\beta_0=10^4$ case, we averaged 
the variables during 200 rotations after the quasi-steady-states are 
achieved. 
For the $\beta_0=10^4$ case, we show the time averages during 25-55 
rotations because the quasi-steady-state is not achieved 
but 90 \% of the total mass escapes at 75 rotations
as a result of the effective mass loss by disk winds.

All the cases of $\beta_0\ge 10^5$ show very similar 
$\langle\beta\rangle$ structures (Figure \ref{fig:betadep}): 
$\beta\approx 100$ at the midplanes, namely the magnetic 
energy can be amplified to $\approx 1$\% of the gas energy. 
The $\beta$ values decrease with increasing height mainly because of 
the decrease of the densities (the magnetic energies stay nearly constant; 
Figure \ref{fig:tav}). 
When the magnetic energy dominates ($\beta\lesssim 1$), the disk 
winds start to accelerate. In the wind regions, the $\beta$ values 
stay at $\beta=0.1$-$1$, rather than further decrease, 
owing to the increased density by the lift-up gas in the winds.         
The locations of the injection regions also concentrate at $|z|/H_0=1.5$-$2$ 
except for the initial $\beta_0=10^{4}$ case. 
The plasma $\beta$ values of the injection regions are 1-10; 
the magnetic energy is comparable to but slightly smaller than 
the equipartition value. 
This condition is favorable for driving mass motions by the breakup 
of large-scale channel flows; 
if the magnetic energy is larger, the field configuration becomes 
more coherent due to the tension so that reconnections hardly occur; 
if the magnetic energy is smaller, 
the reconnections cannot drive strong mass motion.   
The mass flux of the disk winds (Figure \ref{fig:msfxdep}) 
only weakly depends on $\beta_0$ for $\beta_0\ge 10^6$, while it increases 
for smaller $\beta_0$ almost in proportion with $\beta_0$.

\section{Discussions}
We have shown that the gas is lifted up from the injection regions at 
$|z|/H_0\approx 1.5$-$2$ to the surfaces by the Poynting flux and streams out 
from the upper and lower boundaries of the simulation box.  
In effect, 
our results determine the condition of ``mass loading'' in various models of 
global disk wind \citep[e.g.,][Ferreira et al 2006]{bp82,ks97}, 
in which one usually fixes the mass flux in advance 
by setting the densities at the `bases' of winds. 
With our outgoing boundary condition, 
we implicitly assume that once the gas goes out of 
the $z$-boundaries of the simulation box it does not return. 
The validity of this treatment needs to be examined by 
the global modeling of accretion disks. 

Although in the shearing box treatment the time-averaged net mass 
flow in the $x$-direction is zero, 
we can estimate the accretion velocity, $-v_r\approx\alpha c_s^2/r\Omega_0$, 
from the angular momentum balance under steady states, 
where $r$ is a cylindrical distance from a central star \citep{ss73}. 
The ratio of 
the mass-loss rate, $\dot{M}_z$, from the simulation box by the disk winds 
to the mass accretion rate, $\dot{M}_r$, passing through the $y$-$z$ plane 
becomes 
\begin{eqnarray}
\frac{\langle\dot{M}_z\rangle}{\langle\dot{M}_r\rangle}&\approx& 
\frac{\int\int dx dy \langle\rho v_z\rangle}{\int 
dy\langle\Sigma\bar{\alpha}\rangle c_s^2/(r\Omega_0)}=\frac{\langle\langle\rho 
v_z\rangle\rangle{L_x}r\Omega_0}{\langle\langle\Sigma\bar{\alpha}\rangle\rangle 
c_s^2}\nonumber\\ 
&\approx&0.05\left(\frac{r}{10H_0}\right)\left(\frac{L_x}{H_0}\right), 
\label{eq:msfx}
\end{eqnarray}
where $\Sigma=\int dz\rho$ and 
$\bar{\alpha}=\int dz\rho\alpha/\Sigma$;  
$\langle\rangle$ denotes the time-average and $\langle\langle\rangle\rangle$ 
is the average of time and $x$-$y$ planes. 
The final value is estimated from 
$\langle\langle\rho v_z\rangle\rangle\approx 8\times 10^{-5}\rho_{0}c_s$ 
and $\bar{\alpha}\approx 0.012$
in our fiducial run for a moderately thin disk with $H_0/r=0.1$.
We should take {\it the above estimate as an upper limit} because 
the calculation with a larger vertical box size may give a smaller mass flux 
at the top and bottom boundaries.
%
%
The angular momentum loss rates in vertical and radial directions 
give the same scaling,
\begin{equation}
\frac{\langle\dot{\cal{L}}_z\rangle}{\langle\dot{\cal{L}}_r\rangle}=\frac{\int\int 
dx dy\langle\rho v_z\rangle r^2\Omega_0}
{\int dy r\langle\Sigma\bar{\alpha}\rangle c_s^2}=\frac{\langle\langle\rho 
v_z\rangle\rangle L_x r\Omega_0}{\langle\langle\Sigma\bar{\alpha}\rangle\rangle 
c_s^2}\approx\frac{\langle\dot{M}_z\rangle}{\langle\dot{M}_r\rangle},
\end{equation}
because the time-averaged 
 specific angular momentum carried by the winds is approximately the same 
 as that in the disk material at the same radial position 
 in the shearing box treatment. 
In a realistic situation, however, 
 winds possibly carry a larger specific angular 
 momentum than the disk material. 
For such studies, we need to model global accretion disks with disk winds, 
which  
is also important from the viewpoint of 
angular momentum evolution of the star-disk system 
\citep[e.g.][]{mp05}.

Hereafter we discuss the evolution of protoplanetary disks,  
as an application of our results. 
As a reference model, we use the minimum-mass solar nebula (MMSN) of
Hayashi (1981), 
which gives the midplane density, 
$\rho_{0}=1.4\times 10^{-9}\left(\frac{r}{\rm 1AU}\right)^{-11/4}{\rm 
g\;cm^{-3}}$. 
Then, the initial vertical magnetic field of 
$\beta_0=10^6$ in our fiducial run corresponds to $B_{z,0}\approx 0.01$ G, and 
the saturated field strength is $B\approx 1$ G at 1 AU. 

First, we examine how much 
the disk wind contributes to the evaporation of protoplanetary disks 
\citep[see e.g.][for other mechanisms]{dul07}. 
After the saturation of the magnetic fields, 
$\approx 5$\% of the total disk mass is lost from the simulation box from 
200 to 400 rotations by the disk winds in our fiducial case. 
Assuming a disk around a central star 
with the solar mass (1 rotation = 1 yr at 1 AU), we have the timescales 
of the evaporation, $\tau_{\rm ev}\approx 4000$ yr at 1 AU, and 
$6\times 10^5$ yr at 30 AU.
Although this is rather short in comparison with recent 
observational results \citep[typically $\tau_{\rm ev}\sim 10^{6-7}$ yrs, e.g.,
][]{hll01}, 
this is not a severe contradiction because we have not yet taken 
into account the global radial accretion of the disk mass, 
which continuously supplies the mass from the outer region. 
Another important issue that affects, and might reduce, the mass flux of 
disk winds is the effect of resistivity,
which requires an additional detailed analysis of the ionization structure 
\citep{Sano00,InutsukaSano2005} 
and will be the scope of our next paper. 
Here, the estimated $\tau_{\rm ev}$ should be taken as a lower limit. 

The disk scale height has a relation of $H_0/r\propto r^{1/4}$ 
for the MMSN. 
Combining with Equation (\ref{eq:msfx}), we infer that the dynamical 
evaporation by disk winds, in comparison with accretion, 
becomes relatively more important in the 
inner parts of protoplanetary disks than in the outer regions for a constant 
initial $\beta_0$ structure ($B_{z,0}^2\propto r^{-11/4}$ for the MMSN).



Finally, we should point out the effects of waves on dusts 
in protoplanetary disks. 
We have shown that the momentum flux of Alfv\`{e}nic and 
sound-like waves directs to the midplane from the injection regions.  
The momentum flux of the sound-like waves ($\delta\rho\delta v_z$) 
can push dust grains to the midplane by gas-dust collisions. 
Dusts are usually weakly charged; in this case Alfv\'{e}nic waves also 
contribute to the sedimentation of dusts to the midplane through 
ponderomotive force or dust-cyclotron resonance \citep{vj06}.    

This work was supported in part by Grants-in-Aid for 
 Scientific Research from the MEXT of Japan 
 (T.K.S.: 19015004 and 20740100, 
  S.I.: 15740118, 16077202, and 18540238), 
 and Inamori Foundation (T.K.S.). 
Numerical computations were in part performed on Cray XT4 at Center for 
Computational Astrophysics, CfCA, of National Astronomical Observatory 
of Japan. The page charge of this paper is supported by CfCA.

\end{document}